\documentclass[12pt]{iopart}

\usepackage[flushleft]{threeparttable}
\usepackage[utf8]{inputenc}
\usepackage{graphicx}
\begin{document}





\title{Aging in the transport on the corrugated ratchet potential}

\author{Karina I. Mazzitello$^1$, Daniel G Zarlenga$^1$, Fereydoon Family$^2$,  Constancio M Arizmendi$^1$}
\address{$^1$ Instituto de Investigaciones Cient\'{\i}ficas y Tecnol\'ogicas en Electr\'onica, \\
Universidad Nacional de Mar del Plata, B7608 Mar del Plata, Buenos Aires, Argentina}
\address{$^2$ Department of Physics, Emory University, Atlanta, GA 30322, USA} 
\ead{$^1$ kmazzite@mdp.edu.ar}
\vspace{10pt}

\begin{abstract}
Under rapid undercooling, glass forming liquids freeze in an amorphous state that can equilibrate only on enormously long time-scales, This is the characteristic sign of aging, which has been observed in a wide range of systems. Brownian ratchet is a widely studied system that exhibits many types of anomalous dynamical behavior. We have investigated the possibility of aging in the collective motion of Brownian particles in a periodic ratchet potential with quenched disorder. We find that when a slowly growing fraction of particles are trapped for long time, the collective movement tends to become super-diffusive. The super-diffusive transport weakly breaks the ergodicity and the time to cover the whole phase space become enormously long and reminiscent of aging behavior.
\end{abstract}

\vspace{2pc}





\section{Introduction}
\label{Intro}

Disorder in stochastic systems may lead to anomalous behaviors characterized by significant variations of the observed properties, for finite time scales. These variations produce systematic deviations of the dynamical
property mean value which are usually much larger than statistical errors.
A particular class of systems where anomalous behaviors of this type occur are the so-called glass-forming systems.
These systems include, for example, disordered spin systems close to the spin-glass transition \cite{Binder86}, supercooled
molecular liquids \cite{Baran06, Matthai09} and jamming colloidal solutions \cite{Pellet16}. The physics of a glass-forming liquid is such that when
rapidly undercooled under its melting temperature, it loses its ability to flow on experimental time-scales. Glassy materials freeze in an amorphous state that requires enormously long time to equilibrate  \cite{Mezard87,Cavagna09}.
This is the characteristic sign of aging, where systems whose properties depend on the age of the system can be quantified through the so-called relaxation or waiting time. Relaxation times are proportional to the viscosity of the glass and that is why they grow with increasing fluid viscosity.\par

Recently \cite{zia2014, chaudhury2017, SciRep2021Skyrmionratchet}, aging phenomena were studied in several physical systems that involve anomalous dynamics associated with spatial disorder and ratchet potentials. One of these physical systems is a colloidal gel \cite{zia2014}.
In  the coarsening and age-related changes of  an aging colloidal gel, Zia {\it et al.} \cite{zia2014} found that the gel strands contain a glassy, immobile
  interior near random-close packing, surrounded by a liquidlike surface coarsened by the diffusive
  migration of particles. This coarsening is a three-step process: cage
  formation, where particles
  travel rapidly along the network surface until they become bonded to neighbors, cage hopping, characterized by migration of particles between cages, and finally cage trapping where particles get buried within network strands. The motion of the particles is stochastic, but with a net drift to a less-mobile and higher contact number states. Zia {\it et al.} \cite{zia2014} characterize the dynamics as the motion of the particles in a ratchet potential resulting from coarsening and aging. In addition to the work of Zia {\it et al.} \cite{zia2014}, Chadhuri {\it et al.} \cite{chaudhury2017} have made large-scale simulations of aging gels. They also find anomalous behavior in the form of subdiffusive caged dynamics crossing over to large length scale superdiffusive particle motion associated with heterogeneities.\par
  
  Another important example of aging phenomena is exhibited in the dynamics of magnetic skyrmions \cite{SciRep2021Skyrmionratchet} through anomalous behaviors produced by disorder and ratchet motion. Skyrmions are nanometer size, particle-like spin textures present in certain chiral magnets, with enforced stability due to their integer topological charge \cite{SciRep2021Skyrmionratchet}.
  Skyrmions are often considered to be candidates for bits in racetrack data storage devices that may replace typical RAM or HDD memories \cite{SciRep2021Skyrmionratchetref16} in the
near future. The skyrmions advantage over typical ferromagnetic domain walls is because racetrack operates purely electrically and skyrmions can be driven by very low current densities \cite{SCIRep2021Skyrmionratchetref21}.
The mobility of skyrmions is affected by defects in magnetic materials, such as vacancies \cite{PhysRevB100024410(2019)ref.15}, or magnetic grains with varying anisotropy  \cite{PhysRevB100024410(2019)ref.17, PhysRevB100024410(2019)ref.18}. Consequently, skyrmions can undergo pinning due to these magnetic defects. Thus, there exists a depinning threshold for skyrmion motion in the presence of defect induced disorder that leads to different types of flow and transitions between different phases \cite{Reichardt2020ref6, Reichardt2020ref13}. Defects also produce anomalous dynamics and aging in skyrmions 
\cite{PhysRevB97020405(R)(2018), PhysRevB100024410}.\par

In connection with the study of anomalous diffusion, several models have revealed a behavior reminiscent of aging [19, 20].  Khoury et al. [19] found strong fluctuations in the usual ensemble calculation of the diffusion coefficient in the overdamped motion of particles  in a tilted washboard potential, indicating that the system exhibits aging. In [20], the authors have analyzed properties like ergodicity breaking and nonstationarity in the transient motion of inertial particles in a ratchet potential. In this work, we establish a relationship between anomalous diffusion associated with defects, ratchet motion, and aging in a model of corrugated ratchet potential, recently introduced in [20, 21]. Up to our knowledge, this is the first time in which aging is quantified and is related to particle diffusion in a simple ratchet model.

 
 The outline of the paper is as follows. We provide a detailed description of the model and its associated dynamical
 equations, in section \ref{Model}. A brief analysis of the magnitudes of the quenched disorder is presented in the same section.
 The different categories of diffusion characterized by the mean-square displacement and their relations to the mean velocity
 of the particles is discussed in section \ref{transport}. The aging of the transport on the corrugated ratchet potential is studied in
 section \ref{Aging}. This property is revealed by measuring the velocity correlation of the system that depends on two times.
 The same type of dependence is also found in correlation functions of real glasses \cite{Leticia1997, Picco01}. Finally, section \ref{Conclusions} consists
 of concluding remarks.

\section{Model and simulation method}
\label{Model}

The model is defined by considering the overdamped motion of identical non-interacting Brownian particles in a rocking ratchet potential with a quenched disorder \cite{mazzite2021}. The stochastic differential equation (Langevin equation) for such particles is given by 
\begin{equation}
\gamma \dot{x}= F_0 sin(\omega t) + \xi (t) -dU(x)/dx \mbox{,}
\label{Langevin}
\end{equation}
where the left-hand side describes a frictional force experienced by the particles when they move relative to their environment, with $\gamma$ being the drag coefficient. The first term of the right-hand side is an external sinusoidal force with a frequency $\omega$ and amplitude $F_0$ that pushes the particles left and right periodically (rocking ratchet). The second term is Gaussian thermal noise at temperature $T$. The correlation function of the noise obeys
the fluctuation-dissipation relation $\left < \xi (t) \xi (t') \right >= 2\gamma k_B T \delta (t-t')$.
Ultimately, the last term of the right-hand side in Eq. (\ref{Langevin}) is the force due to the ratchet potential $V_p(x)$ and the quenched spatial disorder $V_r(x)$. They contribute to the total potential $U(x)$ through the parameter $\sigma =[0,1]$ according to
\begin{equation}
U(x) = \left ( 1 - \sigma \right ) V_p(x) + \sigma V_r(x) \mbox{.}
\label{total_potential}
\end{equation}
Here $V_p(x)$ is the archetypal ratchet potential of double-sine \cite{Hanggi94}
\begin{equation}
V_p(x)=-V_0 \left [ sin \left ( 2\pi x/\lambda_p\right ) + \left(\mu /2\right) sin \left ( 4\pi x/\lambda_p \right ) \right ]\mbox{,}
\label{ratchet}
\end{equation}
where $V_0$, $\lambda_p$ and $\mu=[0,0.5]$ are the amplitude, the spatial period and the asymmetry, respectively  and $V_r(x)$ is a Gaussian spatially random contribution with correlation function
\begin{equation}
g_r(x)=\left < V_r(x)V_r(0)\right > = g_0 exp \left ( -2\pi^2 x^2/l^2_r \right ) \;\mbox{,}
\label{correlation}
\end{equation}
where $g_0$ is the amplitude and $l_r$ is the ratchet potential roughness. Indeed, smaller values
of $l_r$ lead to a greater number and height of the barriers on the potential $V_p(x)$.\par

\begin{figure}
\centering
\includegraphics[width=30pc]{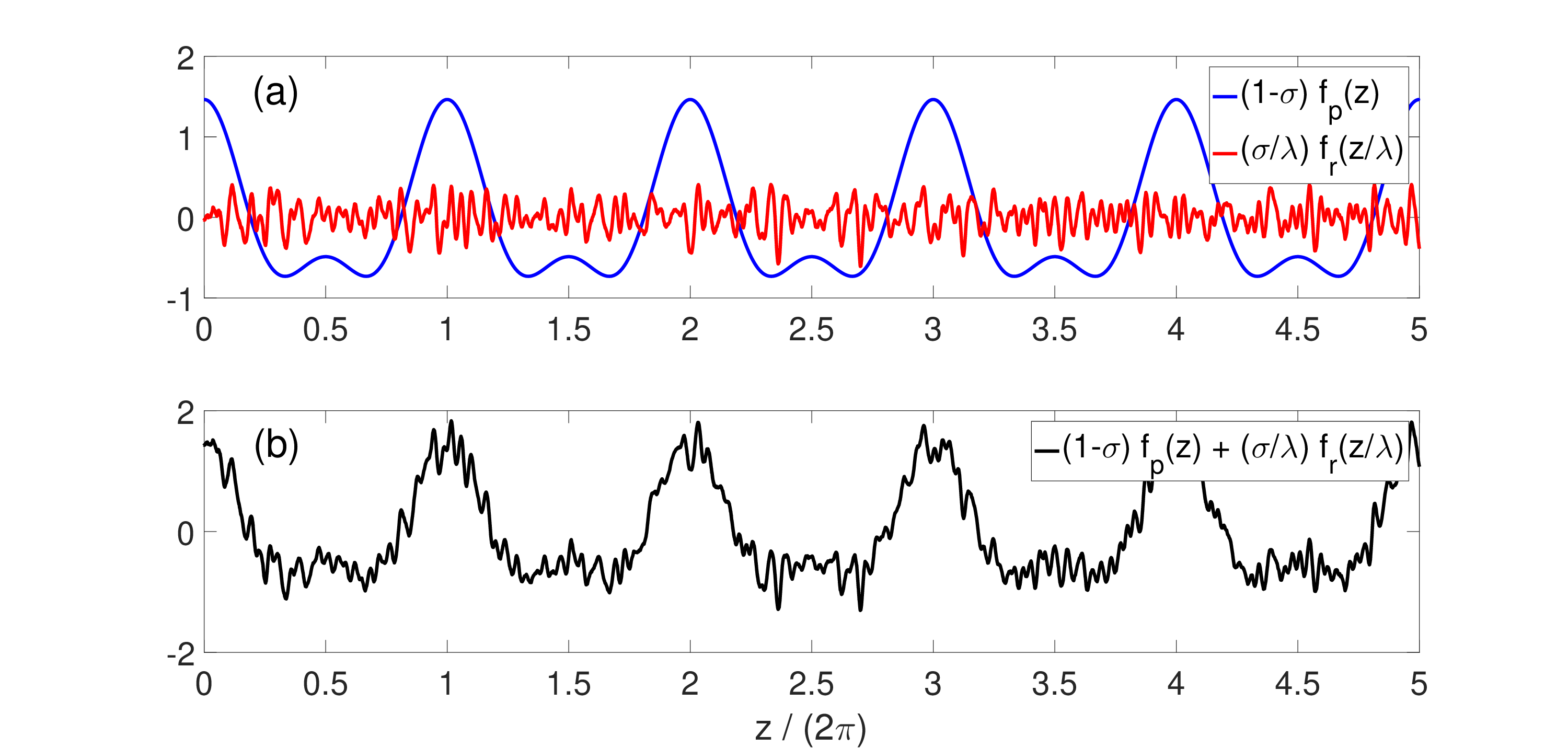}
\caption{\label{U} (a) Relative contributions of a realization of the quenched - disorder force and of the ratchet force, for $\sigma = 0.025$, $\lambda=0.095$, $\mu=0.5$, $\tilde{F_0}=1.47$, $\Omega=0.1$ and $\lambda_p=2\pi$. (b) Resulting force sum of the quenched disorder force and the ratchet force (see Eq. (\ref{Langevin_rescaled})). The disorder level is indeed perceptible for these parameter values.}
\end{figure}

In addition, in order to ensure that the amplitude of the total potential is of the order of $V_0$ independently of $\sigma $, we choose correlations of both ratchet and random potentials to be the same
at $x=0$. The correlation function for the ratchet potential is
\begin{equation}
g_p(x)=V^2_0/2 \left [ cos \left ( 2\pi x/\lambda_p \right ) + \left (\mu /2\right )^2\;cos \left ( 4\pi x/\lambda_p \right ) \right ]
\label{ratchet_correlation}
\end{equation}
Thus, from Eqs. (\ref{correlation}) and (\ref{ratchet_correlation}) and matching $g_r(0)=g_p(0)$, the amplitude of the correlation function (\ref{correlation}) is $g_0=17 V^2_0/32$.\par

The equation of
motion (\ref{Langevin}) can be reduced into a dimensionless form
in terms of the rescaled spatial and temporal quantities $z=2\pi x/\lambda_p $ and $\tau= \left [\left ( 2\pi \right ) ^2 V_0/\gamma \lambda ^2_p\right ] t$ as \cite{Khoury2011}
\begin{equation}
\dot{z}=-(1-\sigma)f_p(z) -\left ( \sigma /\lambda \right) f_r(z/\lambda ) + \tilde{F_0} sin(\Omega \tau ) + \eta (\tau ) \;\mbox{,}
\label{Langevin_rescaled}
\end{equation}
where $f_p$ and $f_r$ are the dimensionless forces arising from the ratchet $V_p$ and random $V_r$ potentials, respectively, 
$\Omega$ and $\tilde{F_0}$ are the frequency and amplitude of the applied dimensionless sinusoidal force, and $\eta (\tau )$ is the dimensionless noise. 
The dimensionless parameters 
\begin{equation}
 \lambda=\frac{l_r}{\lambda_p},\;\; \tilde{F_0}=F_0 \frac{\lambda_p}{2\pi V_0},\;\mbox{and}\; \tilde{T}=\frac{k_B T}{V_0}
 \label{parameters}
 \end{equation}
combined with $\sigma=[0,1]$ and $\Omega=\omega \frac{\gamma \lambda ^2_p}{\left ( 2\pi \right ) ^2 V_0}$ define the 
motion of particles subject to 
Eq. (\ref{Langevin_rescaled}). By using Eqs. (\ref{parameters}), the dimensionless noise correlation function is $\left < \eta(\tau )\eta(\tau ')\right >=2\tilde{T} \delta(\tau -\tau ')$.\par

In Fig. \ref{U}(a), an example of the disorder contributing to the ratchet force is shown, already weighted with the parameters $\sigma = 0.025$ and $\lambda=0.095$, according to Eq. (\ref{Langevin_rescaled}) and for $\mu=0.5$, $\tilde{F_0}=1.47$, $\Omega=0.1$ and $\lambda_p=2\pi$. A decrease of $\lambda$ even
for fixed $\sigma$ leads to an increase in the relative contribution
of the random force. Adding this disorder to the contribution of the ratchet force results in a disorder level perceptible in the scale of the total force (part (b) of Fig. \ref{U}).\par

Throughout this work we set $\sigma=0.025$ and $\lambda=0.095$, associated with a rough total force like that shown in Fig. \ref{U}(b). As we shall see in the next section, increasing $\lambda$ diminishes the effects of the disorder and loses variants of anomalous transport. In addition, 
$\Omega$ and $\lambda_p$ are also set to $0.1$ and $2\pi$, respectively. Variations in $\Omega$ do not lead to any additional phenomenology, while 
the specific choice of $\lambda_p$ is only important as a reference value.\par

We have carried out numerical simulations of the Eq. (\ref{Langevin_rescaled})
over a large number ($10^3-10^4$) of particle trajectories. Each particle moves in a different random potential starting  at  $z=0$. The method to generate the random force values is described in \cite{zarlenga2019} in more detail. 
In order to characterize the transport, we measured the diffusion coefficient $D$ and the mean velocity $<v>$ of the particles, which are given by
\begin{eqnarray}
D(\tau ) &=&\frac{\left < \left ( z(\tau) -\left < z(\tau ) \right >\right )^2 \right >}{2\tau} \mbox{,} \nonumber \\
\left< v(\tau )\right> &=& \frac{\left< z \left(\tau + \Delta \tau \right ) -  z \left(\tau \right )\right>}{\Delta \tau}\mbox{,}
\label{v_MSD} 
\end{eqnarray}
where $\left < ... \right >$ indicates the average taken over the many trajectories and $\Delta \tau$ is chosen equal to $2\pi /\Omega $ (see next section). We also computed the two-time
velocity correlation
\begin{eqnarray}
C_v(\tau, \tau _w)=\frac{ \left < v(\tau )v\left ( \tau_w \right ) \right > - \left < v(\tau )\right > \left < v\left ( \tau _w \right ) \right >}
{\left < v\left ( \tau_w \right )^2 \right > - \left < v\left ( \tau_w \right ) \right >^2\centering}
\label{velocity_correlation}
\end{eqnarray}
to analyze the ergodicity breaking, with $\tau \geq \tau_w$. Note that, the velocity correlation $C_v(\tau, \tau_w)=1$, for $\tau = \tau _w$, and it
decreases as $\tau$ moves away from $\tau_w$. 


\section{Results and Discussion}
\label{Results}

\subsection{Anomalous transport}
\label{transport}

\begin{figure}[htp]
\includegraphics[width=19pc]{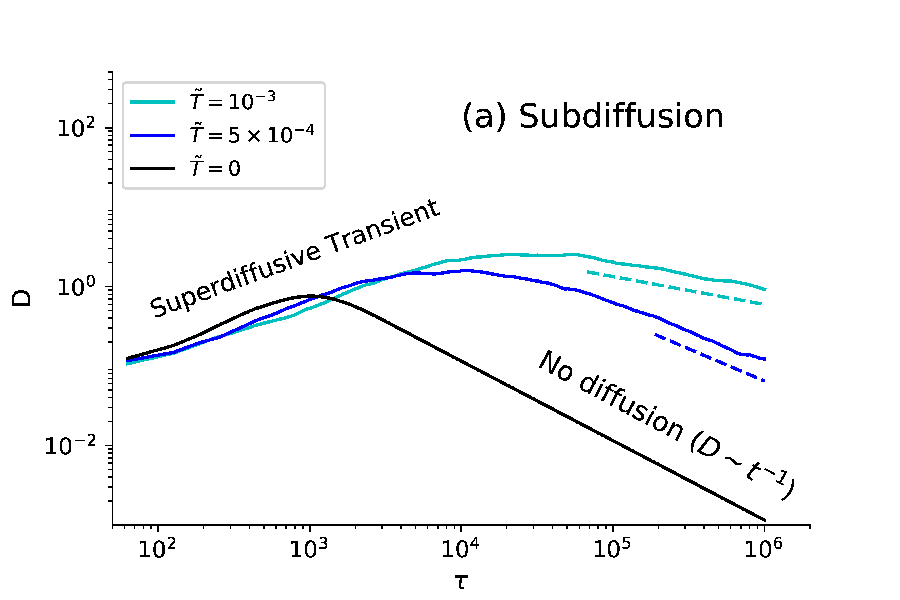}\hspace{2pc}
\includegraphics[width=19pc]{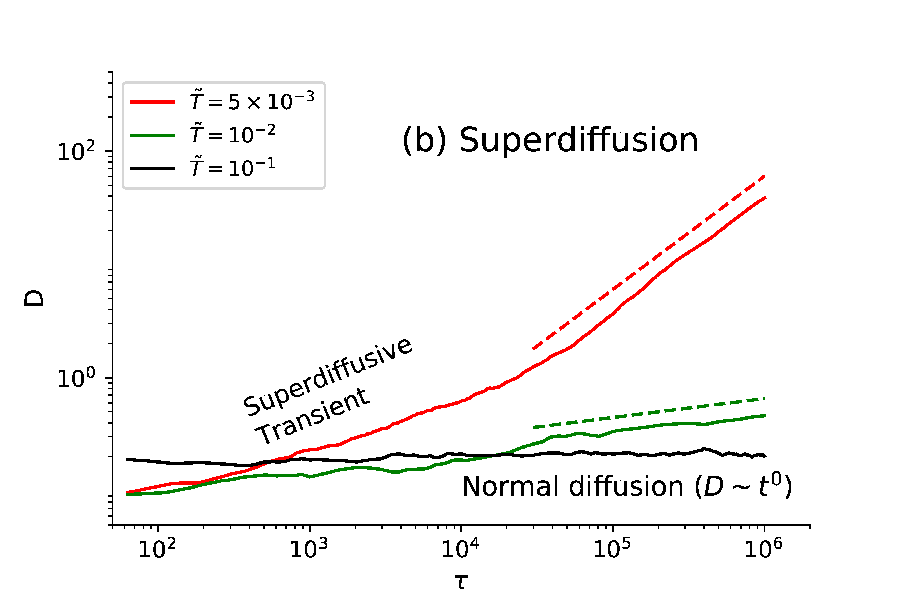}
\caption{\label{C2}Log-log plots of the diffusion coefficient versus the rescaled time,
corresponding to different choices of the dimensionless temperature $\tilde{T}$ as indicated and for fixed external 
force amplitude $\tilde{F_0}=1.47$, quenched noise correlation length $\lambda=0.095$ and ratchet potential asymmetry $\mu =0.1$. (a) Subdiffusive behaviors and a no diffusive case. (b) Superdiffusive behaviors and a diffusive case.  Dashed lines correspond to the asymptotic behavior and their slopes are stored in the fourth column of table \ref{table}.
}
\end{figure}

The quenched noise can have two kinds of effects on diffusion properties: It may only affect the value of the transport coefficients (velocity, diffusion constant, etc.) as compared to the ordered system or it may alter in various ways the diffusion behavior \cite{Bouchad1990}.
We are mainly concerned here with anomalous diffusion phenomena, where the second kind
of effect takes place. The first effect occurs when the external force amplitude is less than the amplitude of the
ratchet in which case, due to the quenched disorder, a transient subdiffusive behavior is found and then the particles
achieve a constant mean velocity with normal diffusion. The second kind of effect is obtained when
the external force amplitude is greater than the amplitude of the
ratchet in which case a variety of anomalous behaviors are obtained. Subdiffusive and superdifusive transports are shown in Fig. \ref{C2}(a) and (b), respectively,
in which $D$ is plotted versus $\tau$ for $\mu=0.1$, $\lambda=0.095$, $\tilde{F_0}=1.47$ at different temperatures. The black curve in part (a), corresponds to  $\tilde{T}=0$, in which, after a  superdiffusive transient all the particles are stopped due to the barriers imposed by the quenched disorder 
\cite{zarlenga2019}. Without thermal noise, the movement has just a transient nature. In contrast, as $\tilde{T}$ increases, anomalous diffusion is reached at long times with $D$ scaling with time as $D(\tau) \sim \tau ^{\beta}$ , where the exponent $\beta$ characterizes the different regimes observed: $-1<\beta < 0$ corresponds to subdiffusion, $\beta > 0$ to superdiffusion and $\beta = 0$ to normal diffusion. The exponent $\beta$ for different temperatures is shown in the fourth column of table \ref{table}.  The motion tends to be subdiffusive at low temperatures, superdiffusive at intermediate temperatures   
and Brownian at high temperatures. The different regimes are related to how  the  particles  explore  the  landscape. As we shall see in section \ref{Aging}, at low temperatures,  most  of  the  particles  explore  every landscape detail and the movement is subdiffusive.  The thermal noise makes the particles jump over many disorder potential peaks and less particles are trapped. In this case, the collective movement  tends  to  disperse  and  it is  superdiffusive. Finally, normal diffusion appears when high enough temperatures dominate particles motion. {\bf Indeed, increasing the temperature, the particles can sort the quenched disorder more easily and thus, the movement tends to disperse less with a higher mean velocity, as will be seen below.}\\
  
\begin{figure}
\centering
\includegraphics[width=22pc]{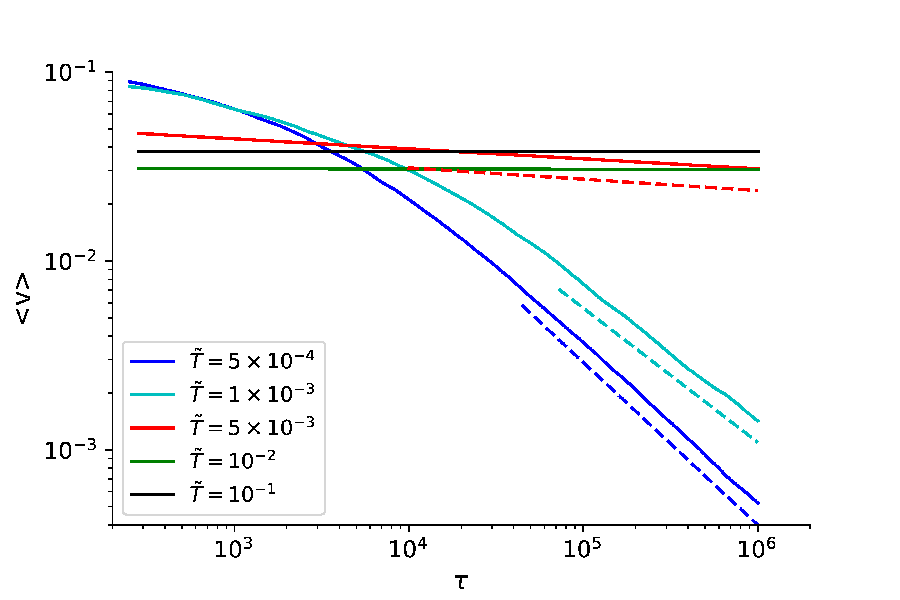}\hspace{2pc}%
\caption{\label{v} A log-log plot of the mean velocity versus the rescaled time, for the same parameters as those corresponding to Fig. 2. Dashed lines correspond to the asymptotic behavior and their slopes are stored in the third column of table \ref{table}. The slopes are equal to zero, for the curves in black and green colors, corresponding to a normal diffusion and a superdiffusive case. In order to minimize the effects of statistical fluctuations, the blue and cyan curves were calculated with $<v>=<z>/\tau$ (see section \ref{transport}). 
}
\end{figure}

The spatial disorder slows down the particle collective motion and the mean velocity decreases over time (subtransport) or at most, it can be constant. This is clearly observed in Fig. \ref{v}, where we show the mean velocity as a function of $\tau$, for the same values of thermal noise as those corresponding to the previous figure. The mean velocity is constant or tends to decrease as $<v> \sim \tau ^{-\alpha }$, with $\alpha >0$ and for $\tau >> 1$.\par

\begin{table}
\centering
\caption{\label{table}
Exponents $\alpha$ and $\beta$ for different values of the
temperature and for fixed external 
force amplitude $\tilde{F_0}=1.47$ and correlation 
length of quenched noise $\lambda=0.095$. $\alpha$ is obtained from asymptotic values of the slope of $log(v)$ versus $log(\tau ) $ (third column see Fig. \ref{v}), while
$\beta$ is obtained in two different ways: from asymptotic values of the slope of $log(D)$ versus $log(\tau ) $ (fourth column see Fig. \ref{C2}) and from Eq.(\ref{exponents}) (fifth column). 
}
\vspace{0.5cm}
\begin{threeparttable}
\begin{tabular}{ccccc}
\hline \hline
\textrm{$\mu $}&
\textrm{$\tilde{T}$}&
\textrm{$\alpha$}&
\multicolumn{1}{c}{\textrm{$\beta$}}&
\textrm{$\beta ^*$}\\
\hline
 & $5\times 10^{-4}$ & $0.86$ & $-0.74$ & $-0.72$\\
 &$10^{-3}$ & $0.71$ & $-0.33$ & $-0.42$\\
 $0.1$ &$5\times 10^{-3}$ & $0.05$ &  $\;\;1.00$ & $\;\;0.90$\\
 & $10^{-2}$ & $0.00$ &  $\;\;0.17$ & ---\\
 & $10^{-1}$ & $0.00$    &  $\;\;1.00$ & ---\\
\hline
 & $3\times 10^{-4}$ & $0.81$ & $-0.66$ & $-0.62$ \\
$0.5$ & & & & \\
 & $10^{-3}$ & $0.00$ &  $\;\;1.27$ & --- \\
\hline \hline 
\end{tabular}
\begin{tablenotes}
\small
\item[*] valid for $\alpha > 0$ (see before Eq.(\ref{exponents}), in the main text).
\end{tablenotes}
\end{threeparttable}
\end{table}

Following the reasoning in \cite{Khoury2011} applied to anomalous diffusion, we establish a relation between the exponents  $\alpha$ and $\beta$ of the velocity and the diffusion coefficient, respectively. As mentioned previously, the quenched noise can alter the very elementary properties of Brownian motion. For Brownian motion, $D$ is a constant given by

\begin{eqnarray}\label{MSD_Normal}
 D(\tau ) = \frac{\lambda_p^2  \left[\left<\tau_p^2\right> - \left<\tau_p\right>^2\right]}{2\left<\tau_p\right>^3}
\end{eqnarray}
provided the first two moments $\left<\tau_p\right>$ and $\left<\tau_p^2\right>$ of the distribution $P(\tau_p)$ are finite, where $\tau_p$ is the time that it takes a single particle to cover the spatial period of the ratchet potential $\lambda_p$ over the course of its trajectory for long enough time. In contrast, when any of the first two moments of $P(\tau_p)$ diverges, anomalous diffusion can be obtained.\par 

According to our numerical results, the asymptotic behavior of the mean velocity can be calculated from Eq. (\ref{v_MSD}), as $<v>=<z>/\tau$ or also as $<v>=\lambda_p/\left<\tau_p\right >$. When the velocity decays as $<v> \sim \tau ^{-\alpha }$, $\left< \tau_p \right >$ varies as $\tau ^{\alpha }$ and therefore the distribution $P(\tau_p)$ has a tail that decreases as a power-law for long stretches of time \cite{Khoury2011}. In this regime, $\left< \tau_p^2 \right > \sim \tau^{\alpha + 1} $ and from Eq. (\ref{MSD_Normal}), we find
\begin{eqnarray}\label{exponents} 
 \beta=1-2\alpha \mbox{,}
\end{eqnarray}
for $\tau \gg 1$. When $0<\alpha<1/2$, the movement of the particles is superdiffusive and  when $1/2<\alpha<1$ the movement is subdiffusive. The values of $\beta$ obtained from Eq. (\ref{exponents}) are listed in the 5th column of Table \ref{table}. These values are consistent with the slopes obtained from Fig. \ref{C2}  (4th column of table \ref{table}) except for $\mu=0.1$ and $\tilde{T}=10^{-3}$ where needs much more time to reach the asymptotic behavior. \par

Finally, when the mean velocity is constant,  $\left< \tau_p \right >=\lambda_p/<v>$ is finite and $\left< \tau_p^2 \right >$ can be finite or diverge over time. In this case, the movement of particles is Brownian or superdiffusive (see the black and the green graphs in Figs. \ref{C2}(b) and \ref{v}).\par

In summary, when $P(\tau_p)$ decreases as a power-law, anomalous diffusion with asymptotic behaviors of $v$ and $D$ is obtained. Moreover, as we shall see in more detail
in the next section, the fat-tailed time distribution promotes aging effects in which most physical properties strongly depend on the history of the sample \cite{Vincent92,Struick78}. The $\tau_p$ correlation is directly related to the spatial correlation, that decays abruptly with  $l_r$ (see Eq. (\ref{correlation})). Therefore,  $\tau_p$ is not correlated because we assume $\lambda_p \gg l_r$, i.e. $\lambda \ll 1$ ( see Eq. (\ref{parameters})). \par  

The effect of the ratchet potential asymmetry on the movement of the particles  plays a similar role as the temperature of the system. Indeed, when the asymmetry $\mu$ increases, superdiffusive behavior is obtained at lower temperatures (see Table \ref{table}, in which exponents for two values of temperature and $\mu=0.5$ are shown for comparison purposes).\par

In order to complete this section, it should be mentioned that, for disorder correlation length much larger than the period of the ratchet potential, that is, negligible randomness, 
normal and superdiffusive behaviors are reached, but the subdiffusive regimes are lost. This has been previously studied in \cite{Khoury2011} 
for particles driven by a constant external force over a landscape consisting of a symmetric periodic potential 
with quenched disorder identical to Eq. (\ref{correlation}).\par

{\bf In brief, the system goes from a subdifusive behavior to a normal diffusion, going before through a superdifussive behavior, when the temperature, the ratchet potential asymmetry or quenched noise correlation length are increased.}


\subsection{Aging in the transport on the corrugated ratchet potential}
\label{Aging}



\begin{figure}
\centering
\includegraphics[width=22pc]{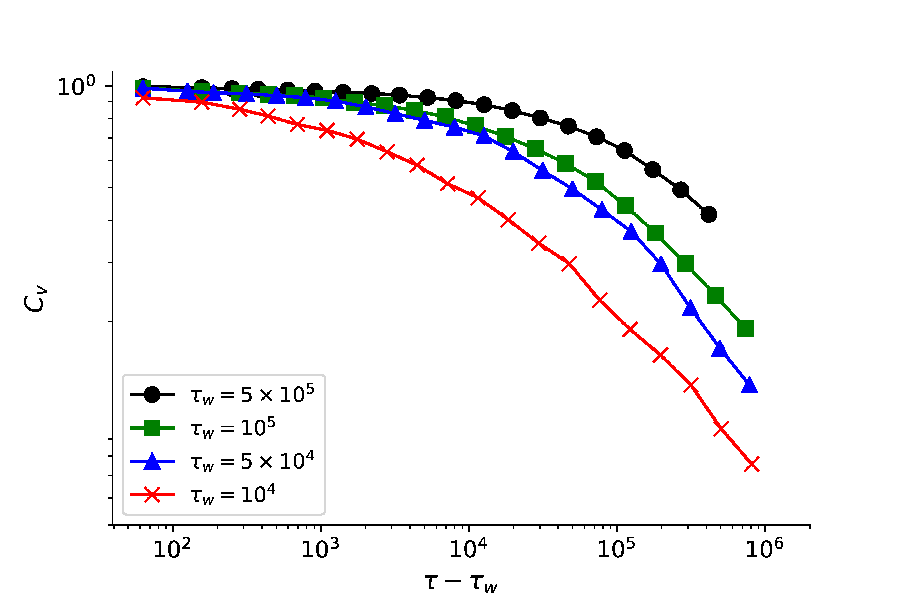}\hspace{2pc}%
\caption{\label{C_v} Two-time velocity correlation function $C_v$ versus $(\tau - \tau_w)$, in log-log scale, corresponding to different choices of $\tau_w$ as indicated and for the superdifussive asymptotic case given by  $\tilde{T}=10^{-3}$, $\tilde{F_0}=1.47$, $\lambda=0.095$ and $\mu =0.5$ (see table \ref{table}). Here $C_v$ was averaged over $10^4$ samples and a temporal statistic average was also done.}
\end{figure}
\begin{figure}
\centering
\includegraphics[width=22pc]{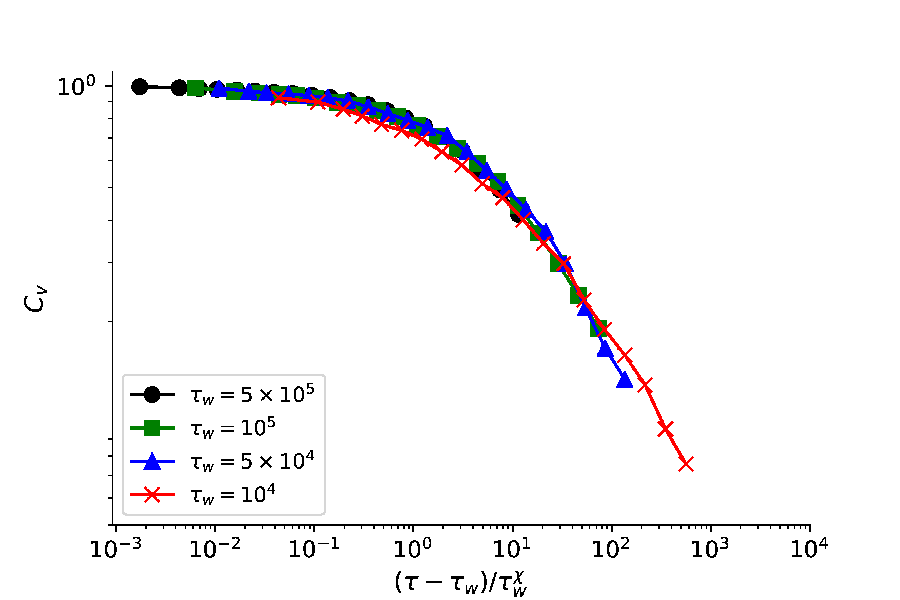}\hspace{2pc}%
\caption{\label{colapso} Scaling of the two-time velocity correlation function $C_v$ versus $(\tau - \tau_w)/\tau^\chi _w$ , obtained by setting $\chi=0.8$, for the data shown in Fig. \ref{C_v}}.
\end{figure}

In equilibrated states and in steady states, physical properties do not change with time. In contrast, in aging systems, the time translation invariance is broken leading to time-dependent dynamics. The aging of the transport on the corrugated ratchet potential can be measured through the two-time
velocity correlation $C_v(\tau,\tau_w)$ given by Eq. (\ref{velocity_correlation}), with $\tau_w$ the observation time or waiting time. Fig. \ref{C_v} shows  $C_v$ as a function of $(\tau - \tau_w)$, corresponding to a asymptotic superdiffusive regime, for different $\tau_w$. The velocity correlation depends on the observation time $\tau_w$, which is the only relevant time scale for the dynamics of this transport. In addition, Fig. \ref{C_v} shows that the older systems, with longer waiting time (circle and square symbols) relax in a slower manner than younger ones (triangle and cross symbols). This can be  quantified, assuming that $C_v$ depends on $\tau/\tau_w^\chi$, with $0<\chi \le 1$. Thus, we propose the scaling on the $x$ axis given by $\left(\tau-\tau_w\right)/\tau_w^\chi$. Figs. \ref{colapso} shows that the data from Fig. \ref{C_v} can be fully collapsed by the previous scaling by taking $\chi=0.8$.\par

{\bf Note that, in this section, we use a ratchet potential asymmetry $\mu=0.5$ instead of $0.1$ due only to a  computational cost that growths with smaller asymmetries. In addition, more samples are required to calculate the velocity correlation function.}

In particular, the cases $\chi<1$ have been called subaging because the effective
relaxation time grows more slowly than the age of the
system \cite{Rinn2000, Bouchaud98}. Indeed, the correlation function $C_v$ is linearly related to the response to an external perturbation through generalized Fluctuation-Dissipation theorem \cite{Sarracino19}. According to this theorem, when the system is disturbed at time $\tau_w$ by a sudden force, the relaxation time is $\tau_w^\chi$. Therefore, the system forgets the perturbation, after a time $\tau_w^\chi$ less than its age $\tau_w$, provided $\chi<1$.\par

\begin{figure}
\centering
\includegraphics[width=22pc]{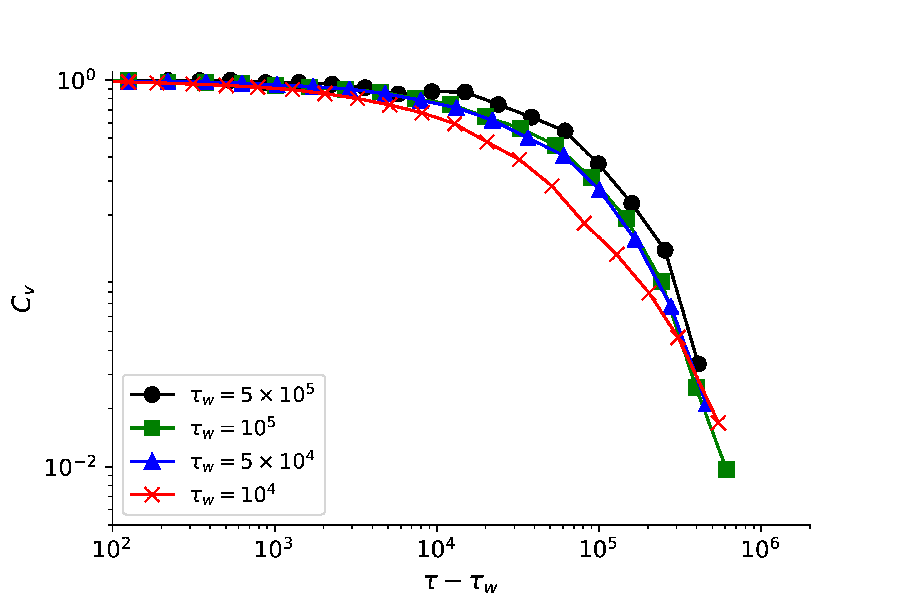}\hspace{2pc}%
\caption{\label{C_v2} Two-time velocity correlation function $C_v$ versus $(\tau - \tau_w)$, in log-log scale, corresponding to different choices of $\tau_w$ as indicated and for the subdifussive asymptotic case given by  $\tilde{T}=3\times 10^{-4}$, $\tilde{F_0}=1.47$, $\lambda=0.095$ and $\mu =0.5$ (see table \ref{table}). Here $C_v$ was averaged over $10^3$ samples and a temporal statistic average was also done.}
\end{figure}

\begin{figure}
\centering
\includegraphics[width=22pc]{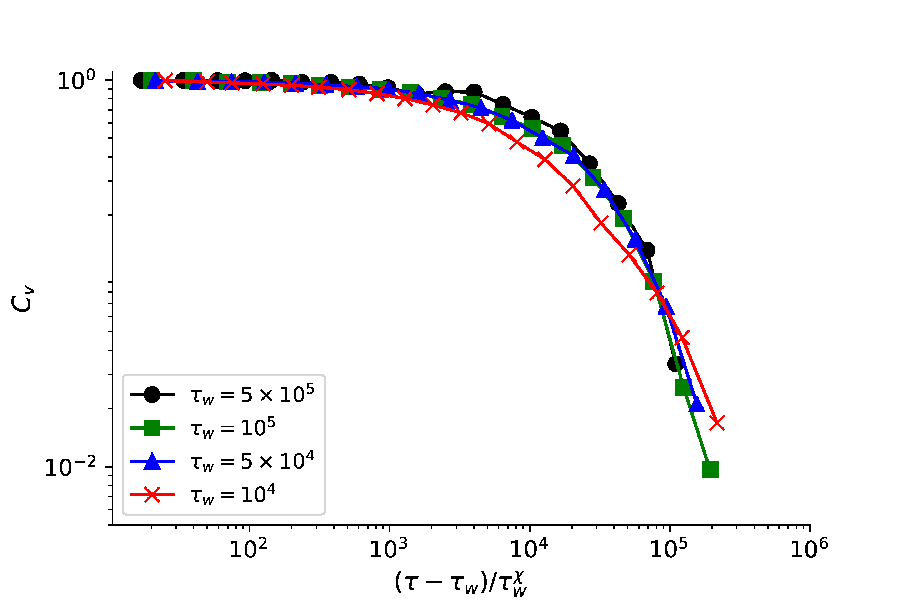}\hspace{2pc}%
\caption{\label{colapso2} Scaling of the two-time velocity correlation function $C_v$ versus $(\tau - \tau_w)/\tau^\chi _w$ , obtained by setting $\chi=0.1$, for the data shown in Fig. \ref{C_v2}.}.
\end{figure}

The aging exponent depends on the way that particles explore the corrugated ratchet potential. Fig. \ref{C_v2} shows the two-time velocity correlation function $C_v$ for an example of subdiffusive transport. Here the aging effect is less than that found for superdiffusive transport. In fact, a good collapse of data of Fig. \ref{C_v2} is obtained in Fig. \ref{colapso2}, in which the x-axis is scaling as $(\tau-\tau_w)/\tau^\chi_w$, with $\chi=0.1$. Regarding scaled figures \ref{colapso} and \ref{colapso2}, we want to point out that the quality of the scaled fits appears to be better for larger values of $\tau_w$.\par


  
\begin{figure}
\centering
\includegraphics[width=18pc]{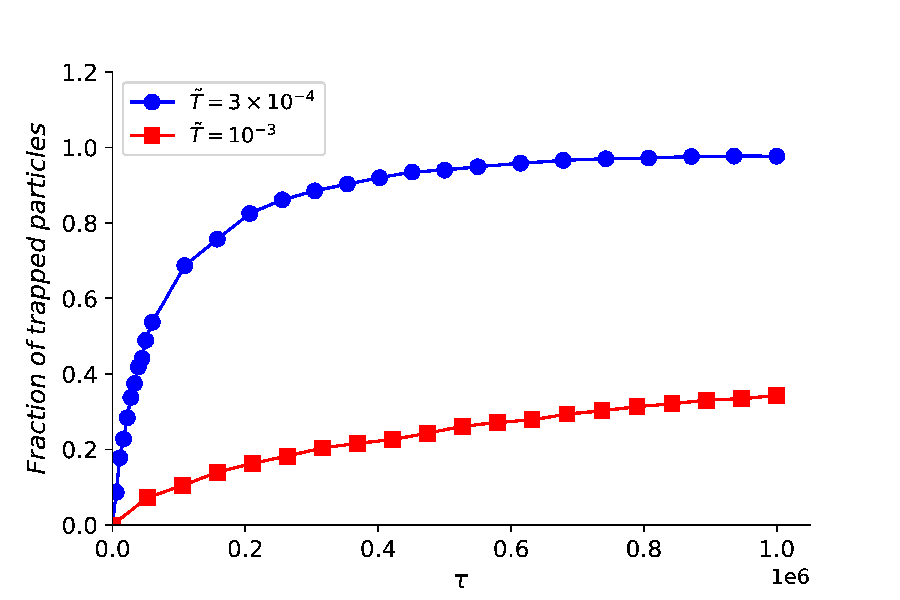}\hspace{2pc}%
\caption{\label{atrapadas} Fraction of trapped particles as a function of $\tau$ in the superdiffusion (red square symbols) and in the subdiffusion (blue circle symbols). The parameters are the same as in Figs. \ref{C_v} and \ref{C_v2}, for the superdiffisive transport and  subdiffusion transport, respectively. The fraction of trapped particles is significantly higher for the subdiffusive movement than for the superdiffusive movement.}.
\end{figure}

Finally, $\chi = 0$ is obtained for normal diffusion. Here $C_v$ depends only on the time difference $(\tau-\tau_w )$ and the system is not aging. In such case, the particles explore the whole phase space because of the presence of thermal fluctuations and the system is ergodic. For $0<\chi \le 1$ the system is also ergodic but exhibits an extremely slow relaxation weakly breaking the ergodicity. For superdiffusive transports,  the number of trapped particles that remain in that state grows slowly with time affecting collective motion. In contrast, for subdiffusive transports, most of the particles explore the traps and the system tends to ergodicity. This is clearly observed in Fig. \ref{atrapadas}, where the fraction of particles trapped in a potential well for a time period $\tau_p$ as a function of $\tau$ are shown for superdiffusive and subdfiffusive transport respectively. Thus, aging related exponent $\chi$ depends on the diffusion of particles on the corrugated ratchet potential.\par 

\section{Conclusions}
\label{Conclusions}
Anomalous diffusion and subaging appear in the motion of particles in a
ratchet potential with both, correlated weak spatial disorder and thermal noise
for different temperatures. The correlation of spatial disorder allows us to
change the rugosity of the ratchet potential. This leads to different categories of diffusion at long times, as the temperature of the system is varied: At high temperatures the particles undergo normal diffusion, whereas at intermediate temperatures we observe a superdiffusive motion and at low temperatures the motion becomes subdiffusive. In all cases, the mean velocity of the particles decreases with time or is at most constant, due to the interplay between the thermal noise and the landscape generated by the spatial disorder.

We find that the mean velocity has a power dependence on time that depends on the type of diffusion. We derived a relation (see Eq. (\ref{exponents})) between the asymptotic exponents of the mean velocity of the particles and those of the diffusion coefficient. In addition, we find that the system exhibits a behavior reminiscent to subaging and weak nonergodicity. When the details of the landscape do not affect the particle motion and a slowly growing fraction of particles remain trapped for long time, the dynamics becomes superdiffusive. This breaks the ergodicity because the time for the system to explore the whole phase space becomes enormously long
\cite{Spiechowicz2016}. We find subaging in superdiffusive transport. On the other hand, the system tends to recover its ergodicity in the subdiffusive case. These behavior were found after a superdiffusive transient due to a high quenched disorder and for an amplitude of the external force that is higher than the amplitude of the ratchet force. In contrast, a regular ratchet movement is found for an amplitude of the external force that is less than the amplitude of the ratchet force after a subdiffusive transient. In this case, a net motion is obtained for high enough temperatures that combined with the external force let the particle overcome both the ratchet potential and the quenched disorder.\par

In summary, we found a set of dramatic anomalous behaviors as diverse as subtransport, subdiffusion, and superdiffusion with subaging when the periodic ratchet potential is modified with a small amount of correlated weak disorder and thermal noise.

\section{Bibliography}
\bibliographystyle{elsarticle-num}

\bibliography{ratchet_aging.bib}







\end{document}